# Modeling and Analysis of Heat Transfer and Fluid Flow Mechanisms in Nanofluid Filled Enclosures Irradiated from Below

Inderpreet Singh[a], Satbir Singh Sehgal[a], and Vikrant Khullar[b,*]

[a]Mechanical Engineering Department, Chandigarh University, Gharuan, Mohali-140413, Punjab, India
[b]Mechanical Engineering Department, Thapar Institute of Engineering and Technology, Patiala-147004, Punjab, India

*Corresponding author. Email address: vikrant.khullar@thapar.edu

**ABSTRACT**
Radiation-driven transport mechanisms are ubiquitous in many natural flows and industrial processes. To mimic and to better understand these processes, recently, radiatively heated nanofluid filled enclosures have been extensively researched. The present work is essentially a determining step in quantifying and understanding the transport mechanisms involved in such enclosures. In particular, a two-dimensional square nanofluid filled enclosure irradiated from the bottom has been investigated in laminar flow situation. Effects of nanofluid optical depth, inclination angle of the enclosure, incident flux, and boundary conditions (adiabatic and isothermal) have been investigated. Moreover, the temperature and flow fields have been carefully analyzed in the situation ranging from "volumetric" to "mixed" to "surface" absorption modes. Under adiabatic boundary conditions, steady state is unconditionally achieved irrespective of the incident flux magnitude (varied between 5Wm$^{-2}$ - 50Wm$^{-2}$), enclosure inclination angle (varied between 0° to 60°) and mode of absorption (surface, mixed or volumetric). However, in case of isothermal boundaries; onset of natural convection and its transition into transient regime is significantly impacted by the mode of absorption and the enclosure inclination angle.

## 1. INTRODUCTION

Radiatively heated nanofluid filled enclosures lend themselves to be employed for understanding numerous human-engineered (heating and cooling in nuclear reactors, electronic components, and solar thermal systems, etc.) and natural (viz., flows in Earth's mantle and frozen lakes, etc.) processes.  These enclosures are broadly categorized based on the mode of absorption mechanism operational; viz., surface or volumetric absorption. Surface absorption is the classical supposition followed extensively in numerous radiation-driven flows, where convective cells (Benard cells) are generated by maintaining a fixed temperature difference between confined cold top and hot bottom surface of the nanofluid filled enclosure. Surface heating at the bottom results in Rayleigh Benard convection (RBC); this can be extended to volumetric absorption mode when radiative flux is allowed to interact with bulk of the fluid.
In literature, a lot of work has been reported in relation to nanofluid filled enclosures undergoing surface heating - i.e. nanofluid undergoing RBC. In particular, two-dimensional enclosures (of various aspect ratios packed with different nanofluids (and varying nanoparticles volume fractions) with differential heated walls have been extensively studied both experimentally and mathematically. The key conclusions being that dispersion of nanoparticles results in enhancement in heat transfer rate at optimum nanoparticle volume fractions, aspect ratio, and Rayleigh number (Khanfer et. al.,2003 [M]; Ho et. al., 2010 [E]; Hwang et. al., 2007 [M]; Corcione, 2010 [M]; Garbadeen et. al. 2017 [E]; Joshi and Pattamatta, 2017 [E]; Sharifpur et. al., 2018 [E]; Giwa et. al.,2020 [E]). Furthermore, researchers have also investigated the effect of enclosure inclination angle on temperature and flow fields (Abu-Nada, 2009[M], Kahveci, 2010[M], Bouhalleb et. al., 2014 [M], Torki and Etesami, 2019 [E]).

However, there are only a few studies in the literature that report bottom-irradiated volumetric heating of nanofluid. These studies point out that better photo-thermal conversion could be achieved for bottom irradiated nanofluid filled enclosures in comparison to top irradiated nanofluid filled enclosures (Wang et. al., 2019; Wang et. al., 2020a [E]; Wang et. al., 2020b [E]; Wang et. al., 2021 [E]). Aforementioned reported works are limited to turbulent flow situation and that too for very specific boundary conditions (see Fig. 1). Through present work, we have developed a comprehensive theoretical framework to cover host of nanoparticle materials (by varying absorption coefficient) and operating parameters (boundary conditions and inclination angle). Overall, the objective is to decipher transport mechanisms operational in bottom irradiated nanofluid filled enclosures in laminar flow situation.

Fig. 1 Selected works pertinent to radiation-driven nanofluid filled enclosures irradiated from the bottom.

## 2. BASIC CONSTRUCTIONAL DETAILS OF NANOFLUID FILLED ENCLOSURES

Figure 2 represents the schematic diagrams of two configurations for bottom irradiated nanofluid filled enclosures studied in the present work. In particular, nanofluid filled enclosures with adiabatic and isothermal boundaries (which define the extreme limits of energy storage and energy discharging capacities respectively) have been investigated. To keep the analysis general, instead of varying nanoparticle material/volume fraction; absorption coefficient has been varied. Very large and small values of absorption coefficient simulate surface absorption mode (energy absorption being confined to bottom and top surfaces respectively); whereas, moderate values simulate volumetric absorption mode. Furthermore, an attempt has been made to analyze the effect of enclosure inclination angle and incident flux on the momentum and energy transport mechanisms.

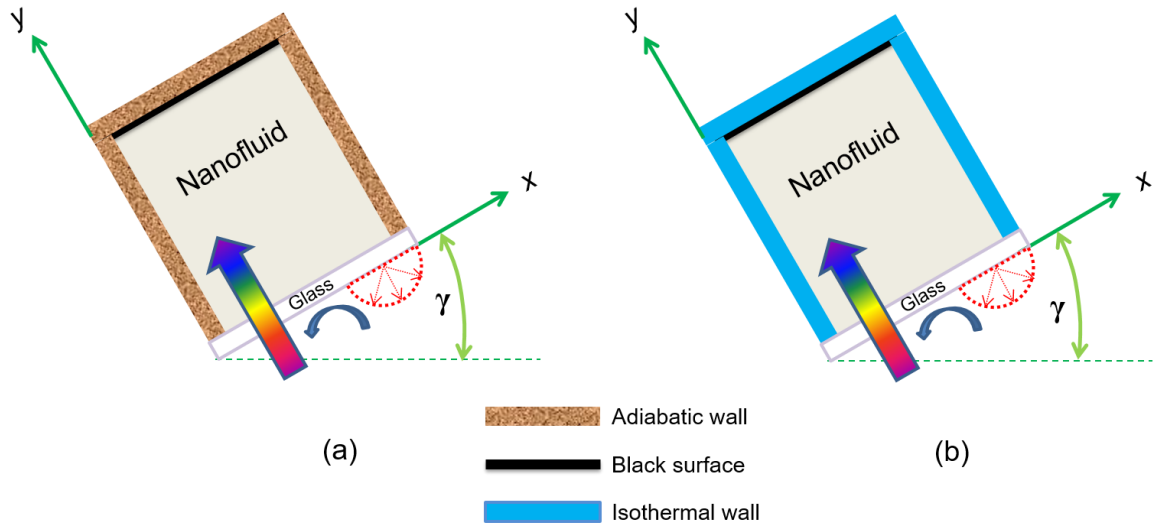

Fig. 2 Schematic representation of bottom irradiated nanofluid filled enclosure with (a) adiabatic boundaries and (b) isothermal boundaries

## 3. THEORETICAL MODELING FRAMEWORK

### 3.1 Underlying assumptions
- Top wall of the enclosure is considered to be perfectly black; i.e., it absorbs all the radiation that is able to reach the top surface.
- Given the fact that the basefluid is highly absorbing in infrared region "in-scattering" and "emission" within the working nanofluid has been neglected (Lenert and Wang 2012).
- Incident flux is assumed to strike normally at the bottom surface.
- Transmissivity (to the incident radiation) and emissivity (in the infrared region) of glass cover have been assumed to be unity.
- The nanofluid is assumed to be stable; therefore, "single-phase model" is used for investigating nanofluid filled enclosures (Singh et. al., 2020).
- Nanoparticles volume fractions being relatively low, thermophysical properties for nanofluid have been assumed to have values similar to that of the basefluid (Singh et. al., 2020)

### 3.2 Mathematical modeling of transport phenomena for predicting flow and temperature fields
To model and analyze the transport mechanisms for bottom irradiated nanofluid filled enclosures (both in surface and volumetric absorption modes); mass, momentum, and energy equations have been employed.

$$\frac{\partial(\rho u_{nf})}{\partial x} + \frac{\partial(\rho v_{nf})}{\partial y} = 0 \qquad (1)$$

$$\frac{\partial(\rho u_{nf})}{\partial t} + \frac{\partial(\rho u_{nf} u_{nf})}{\partial x} + \frac{\partial(\rho v_{nf} u_{nf})}{\partial y} = -\frac{\partial p_{nf}}{\partial x} + \mu_{visc}\left(\frac{\partial^2 u_{nf}}{\partial x^2} + \frac{\partial^2 u_{nf}}{\partial y^2}\right) + S_{bx} \qquad (2)$$

$$\frac{\partial(\rho v_{nf})}{\partial t} + \frac{\partial(\rho u_{nf} v_{nf})}{\partial x} + \frac{\partial(\rho v_{nf} v_{nf})}{\partial y} = -\frac{\partial p_{nf}}{\partial y} + \mu_{visc}\left(\frac{\partial^2 v_{nf}}{\partial x^2} + \frac{\partial^2 v_{nf}}{\partial y^2}\right) + S_{by} \qquad (3)$$

where $u_{nf}$ and $v_{nf}$ are the nanofluid velocities in $x$ and $y$ direction; $\mu_{visc}$ is the dynamic viscosity of working nanofluid; $p_{nf}$ is effective pressure; $S_{bx}$ and $S_{by}$ represent the Boussinesq source terms in $x$ and $y$ directions respectively.

$$S_{bx} = \rho_{ref} \beta g \sin(\gamma)(T_{nf} - T_{ref}) \qquad (4)$$

$$S_{by} = \rho_{ref} \beta g \cos(\gamma)(T_{nf} - T_{ref}) \qquad (5)$$

where $\beta$ is the coefficient of thermal expansion, $\rho_{ref}$ is the reference density of nanofluid, $g$ is the acceleration due to gravity, $T_{nf}$ is the local temperature of nanofluid and $T_{ref}$ is the reference temperature.

$$\frac{\partial(\rho T_{nf})}{\partial t} + \frac{\partial(\rho u_{nf} T_{nf})}{\partial x} + \frac{\partial(\rho v_{nf} T_{nf})}{\partial y} = \frac{k_{nf}}{C_{ps,nf}}\left(\frac{\partial^2 T_{nf}}{\partial x^2} + \frac{\partial^2 T_{nf}}{\partial y^2}\right) + \frac{S^T}{C_{ps,nf}} \qquad (6)$$

where $S^T$ is the radiative source term representing flux decay within the physical domain of enclosures and expressed as Eq. (7)

$$S^T = -\frac{\partial q_r}{\partial y} \qquad (7)$$

where $q_r$ represents the radiative flux.

The preceding governing differential equations are subjected to the following boundary conditions:

$$u_{nf} = v_{nf} = 0, \text{ for } x = 0 \text{ and } x = D \qquad (8a)$$

$$u_{nf} = v_{nf} = 0, \text{ for } y = 0 \text{ and } y = H \qquad (8b)$$

$$T_{nf}(y,0) = T_i = T_{amb} \qquad (8c)$$

At the bottom surface, convective and radiative losses are calculated using Eq. (9)

$$-k_B a_B \frac{\partial T}{\partial y}\bigg|_{y=0} = h_B a_B (T_s - T_{amb}) + \sigma a_B (T_s^4 - T_{amb}^4) = Q''_{loss\_conv} + Q''_{loss\_rad} \qquad (9)$$

where, $h_B = 0.27 Ra^{1/4} ........ 10^5 \leq Ra \leq 10^{10}, \Pr = 0.71$ (10a)

$h_r = \sigma(T_s^2 + T_{amb}^2)(T_s + T_{amb})$ (10b)

At the top, left and right boundaries two cases have been considered

Case 1: Adiabatic boundaries

$-k_T a_T \dfrac{\partial T}{\partial y}\Big|_{y=H} = h_T a_T (T_T - T_{amb})$ (11a)

$-k_L a_L \dfrac{\partial T}{\partial y}\Big|_{x=0} = h_L a_L (T_L - T_{amb})$ (11b)

$-k_R a_R \dfrac{\partial T}{\partial y}\Big|_{x=L} = h_R a_R (T_R - T_{amb})$ (11c)

where convective heat transfer coefficients corresponding to left, right and top walls of enclosure are $h_L = h_R = h_T = 0$ Wm$^{-2}$K$^{-1}$.

Case 2: Isothermal boundaries

$T_{nf}(y, H) = T_{amb}$ (12a)

$T_{nf}(x, 0) = T_{amb}$ (12b)

$T_{nf}(x, L) = T_{amb}$ (12c)

Furthermore, to calculate flow patterns, stream function values are obtained through Eqs. (13a) and (13b) as

$\dfrac{\partial \Psi_{nf}}{\partial x} = -v_{nf}$ (13a)

$\dfrac{\partial \Psi_{nf}}{\partial y} = u_{nf}$ (13b)

### 3.3 Numerical modeling

Governing differential equations (pertinent to conservation of mass, momentum, and energy) are numerically solved using finite control volume approach. SIMPLE (semi-implicit method for pressure-linked equations) algorithm has been applied to momentum equations that are linked through pressure term in a non-linear manner. Figure 3 details the computational algorithm and typical properties/parameters employed. The convective part of the momentum and energy equations follows the first-order upwind scheme while the diffusive part follows the central difference scheme (Versteeg et al., 2007 & Özisik et al., 2017). Meanwhile, SOR iterative scheme is applied to obtain a converging solution.

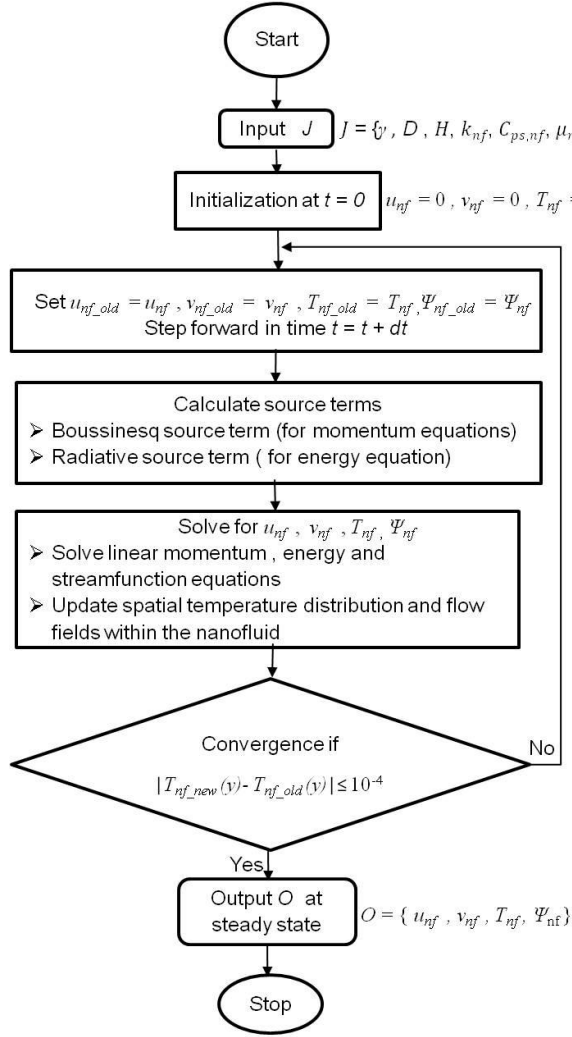

(a) (b)

Fig. 3: (a) Algorithm implemented in MATLAB to determine the flow and temperature field, and (b) typical values of properties/parameters employed.

### 3.4 Grid independence test

To identify the optimum number of control volumes for domain discretization so that the results are independent of the grid size; mid-plane temperature along the enclosure height has been computed for various grid sizes. Mid-plane temperatures for grid sizes of $60 \times 60$, $80 \times 80$, $100 \times 100$, $120 \times 120$, $160 \times 160$ and $180 \times 180$ are reported in Fig.4. Reported results show the mid-plane temperatures for the grid sizes viz., $120 \times 120$, $160 \times 160$, and $180 \times 180$ nearly overlap. Therefore, in the present work $120 \times 120$ grid size has been chosen to determine the flow and temperature fields.

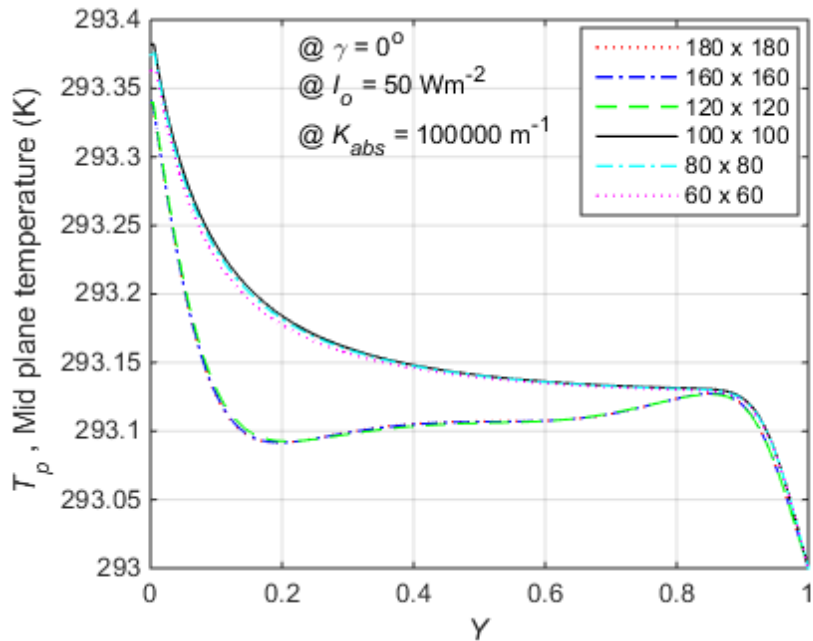
Fig. 4 Mid-plane temperature for various grid sizes.

### 3.5 Model Validation
The developed model has been validated with numerical results presented in Deshmukh et al. 2011. Herein, an inclined square cavity (with all the four walls being isothermal and maintained at 0K) housing working fluid ($Pr = 0.71$) with uniform heat generation has been investigated (see Fig. 5). Maximum dimensionless temperature for Rayleigh number ranging from 1 to $1.5 \times 10^5$ and corresponding to inclination angles $15^o$, $30^o$, and $45^o$ have been reported.

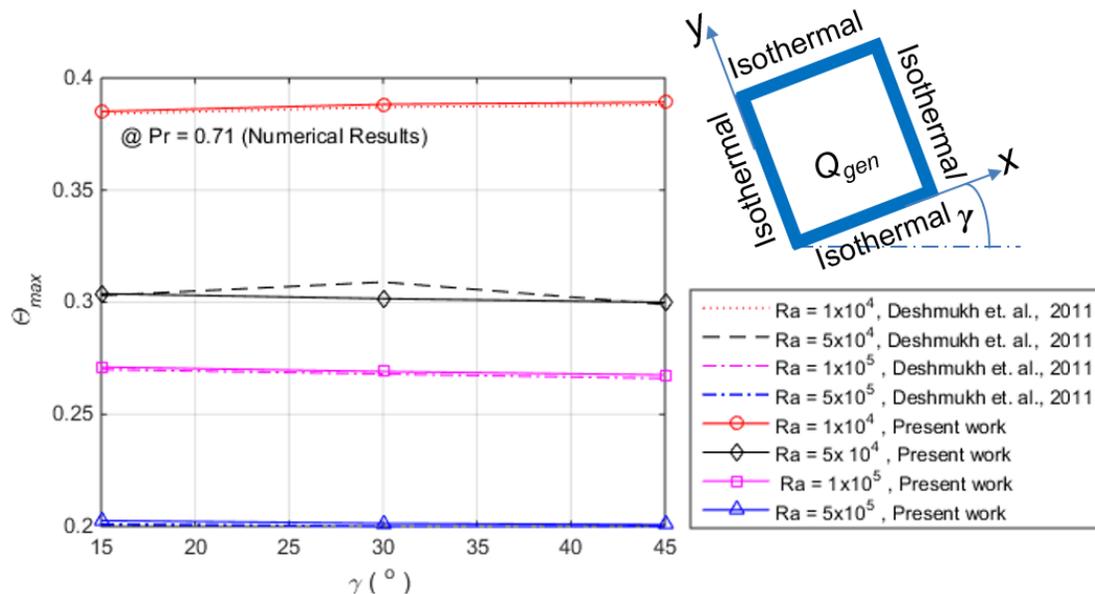
Fig. 5 Validation of the developed model with Deshmukh et. al., 2011: "maximum dimensionless temperature" as a function of enclosure inclination angle for various Rayleigh numbers.

## 4. RESULTS AND DISCUSSION

### 4.1 Effect of absorption coefficient on the incident flux attenuation along the enclosure height

The magnitude of absorption coefficient dictates the distribution of energy generation term (i.e., the divergence of the radiative flux) within the nanofluid. Figure 6 shows the values of dimensionless flux as a function of dimensionless enclosure height. Clearly, 'very high' value of absorption coefficient results in the entire energy getting absorbed at the bottom surface - representing typical 'surface absorption mode'. On the other hand 'low to moderate' values of absorption coefficient results in 'mixed to typical volumetric' absorption modes respectively.

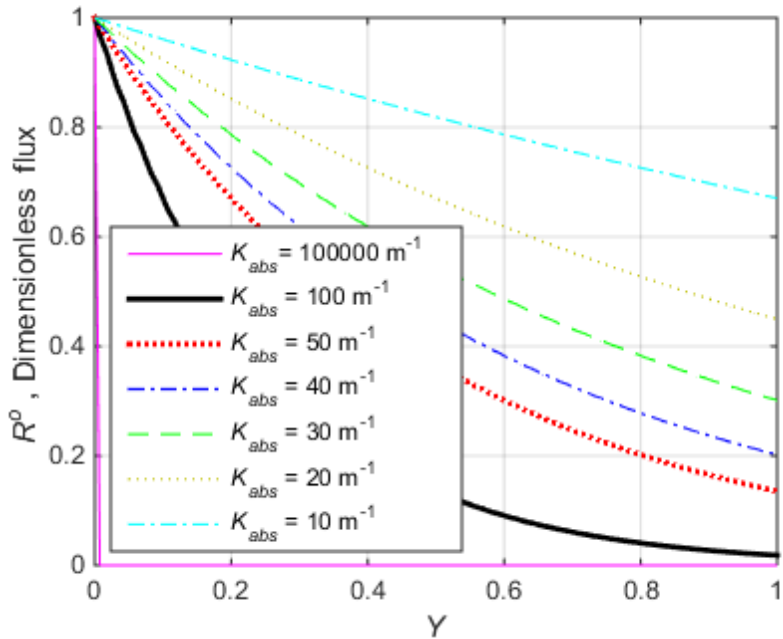

Fig. 6 Flux attenuation along with enclosure height corresponding to, $I_o$ = 50 Wm$^{-2}$ under different absorption coefficients approximating volumetric and surface absorption modes.

### 4.2 Assessing the key transport mechanisms in 'surface' and 'mixed to volumetric' absorption modes in nanofluid filled enclosure with adiabatic boundaries

#### 4.2.1 'Mixed to volumetric' absorption mode (adiabatic boundaries)

*Effect of absorption coefficient:* Figure 7 shows temperature distribution for 'mixed to volumetric' absorption mode corresponding to enclosure inclination, $\gamma = 0^o$. Results show horizontal isotherms for all absorption coefficients along enclosure height - signifying conduction as the dominant mode with negligible buoyancy effects. Moreover, as absorption coefficient, $K_{abs}$ approaches larger values ($K_{abs}$ = 100 m$^{-1}$), the maximum temperature obtained, starts dispersing within a broader region of the enclosure. This may be ascribed to the fact that the radiation source term synergizes the diffusion mechanism more pronouncedly at higher absorption coefficients (clearly apparent from the slopes of mid-plane temperatures in Fig. 8(a)).

Furthermore, it may be noted that similar temperature field is also observed for higher values of incident flux.

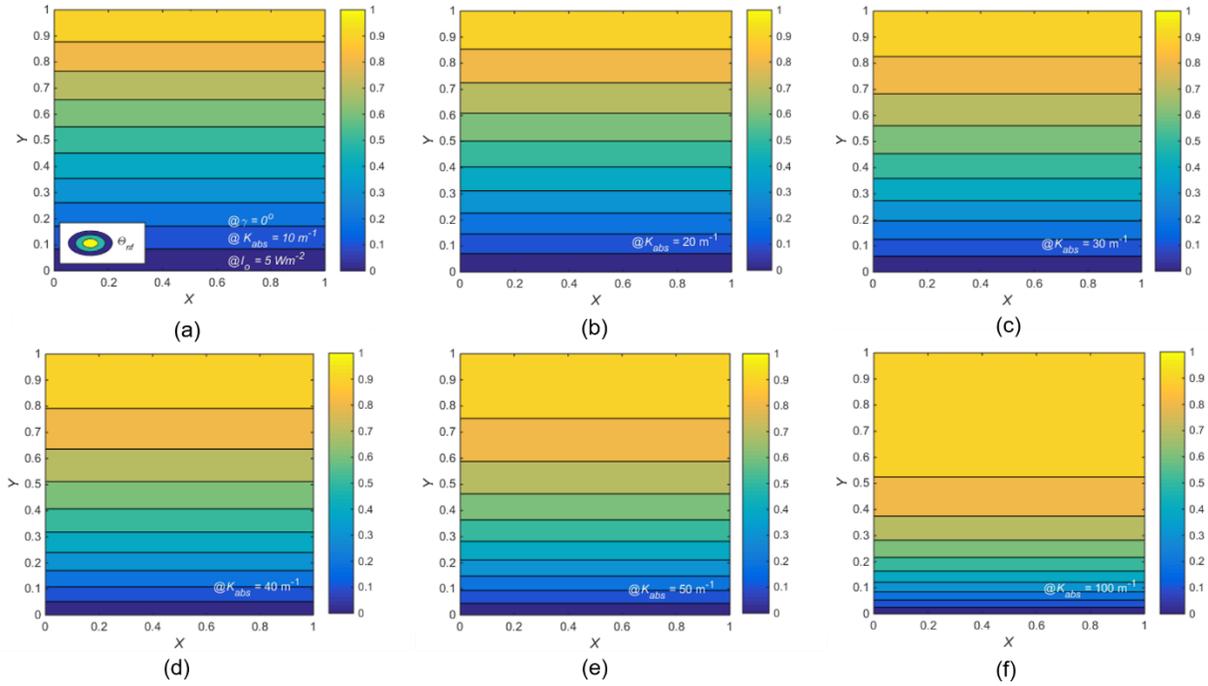

Fig. 7 Temperature contours for 'mixed to volumetric' absorption mode for nanofluid filled enclosures at $\gamma = 0°$, $I_o = 5$ Wm$^{-2}$; nanofluid absorption coefficient (a) $K_{abs} = 10$ m$^{-1}$ (b) $K_{abs} = 20$ m$^{-1}$ (c) $K_{abs} = 30$ m$^{-1}$ (d) $K_{abs} = 40$ m$^{-1}$ (e) $K_{abs} = 50$ m$^{-1}$ (f) $K_{abs} = 100$ m$^{-1}$

*Effect of enclosure inclination angle:* In the foregoing discussion, radiation and conduction have been found to be as the predominant modes of heat transfer for nanofluid filled enclosure at $\gamma = 0°$. However, as the inclination angle increases to higher values ($\gamma = 30°, 45°, 60°$), the profiles for mid-plane temperature show more uniformity (see Figs. 8(b), 8(c), and 8(d)). These variations in mid-plane temperature profiles (relative to the case of $\gamma = 0°$) may be attributed to the onset of convection in addition to the conduction and radiation (heat generation). Similar trends are observed even for higher values of flux (for instance for $I_o = 50$ Wm$^{-2}$) with the only difference that at high flux values convection mode becomes more pronounced (see Figs. 9 (a), 9(b), 9(c) and 9(d)).

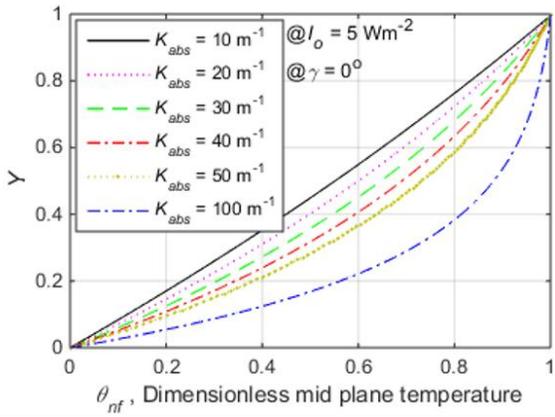
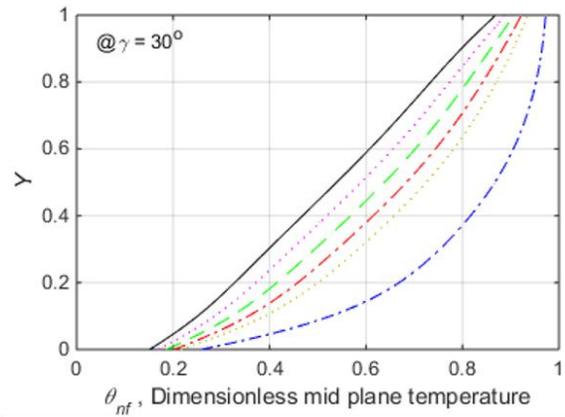
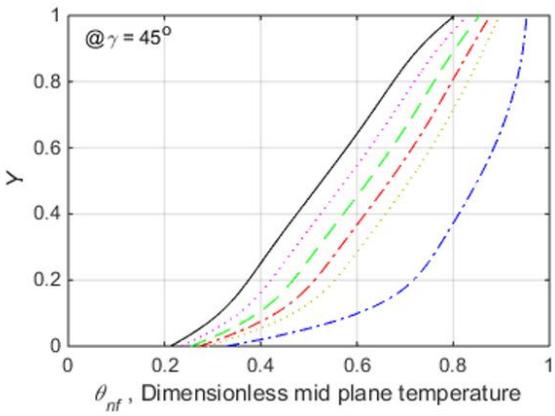
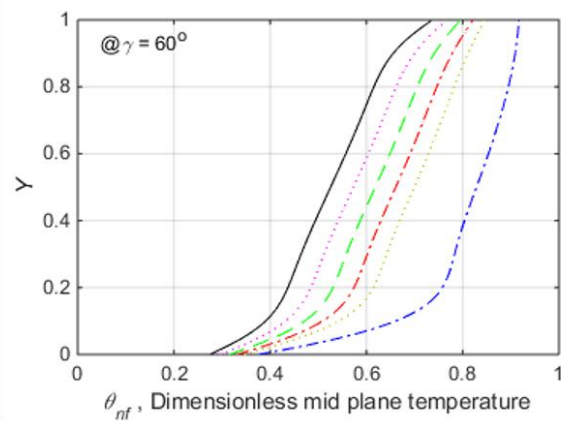

Fig. 8 Mid-plane temperature profiles for 'mixed to volumetric' absorption mode for nanofluid filled enclosures at, $I_o = 5$ Wm$^{-2}$ and inclination angles of (a) $\gamma = 0°$, (b) $\gamma = 30°$, (c) $\gamma = 45°$, and (d) $\gamma = 60°$.

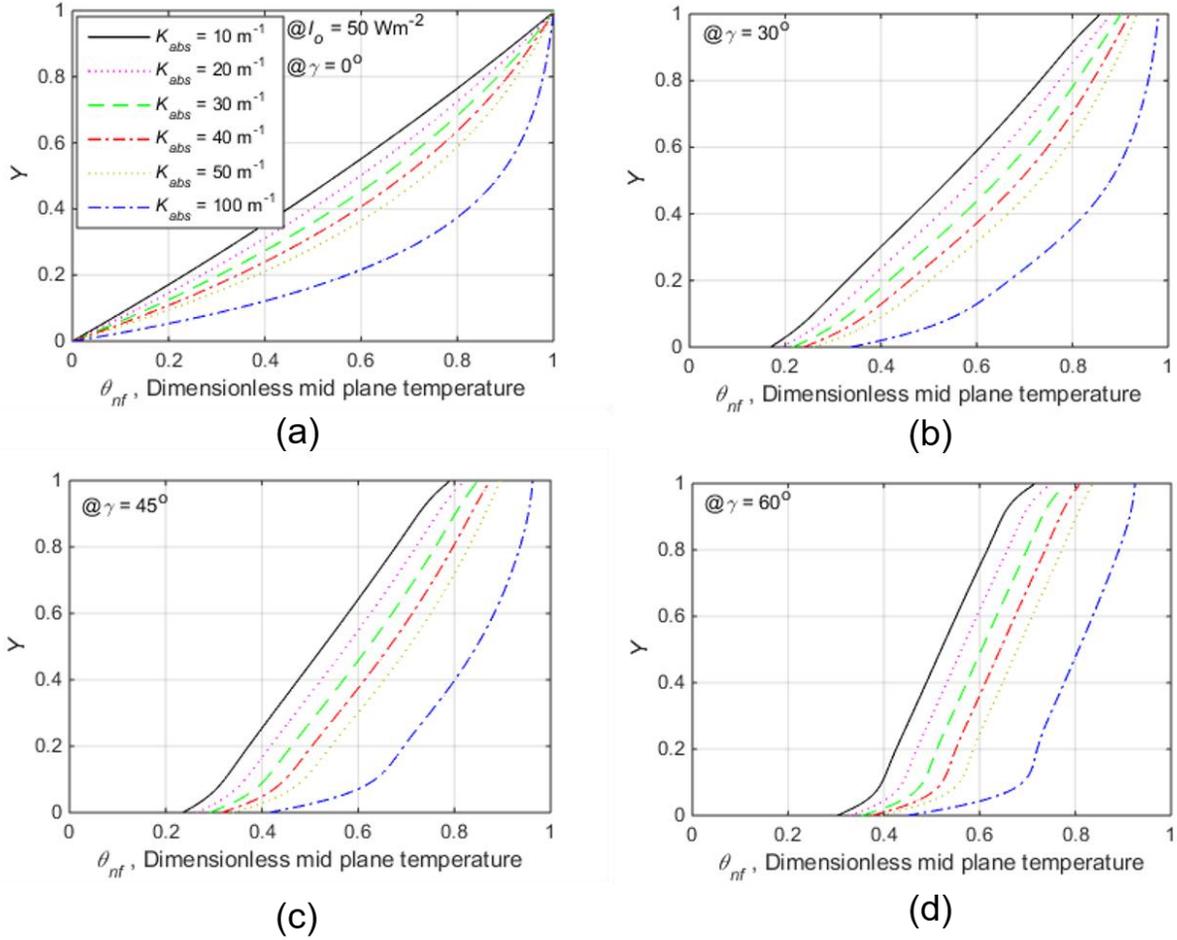

Fig. 9 Mid-plane temperature profiles for 'mixed to volumetric' absorption mode for nanofluid filled enclosures at, $I_o = 50$ Wm$^{-2}$ and inclination angles of (a) $\gamma = 0°$, (b) $\gamma = 30°$, (c) $\gamma = 45°$, and (d) $\gamma = 60°$.

Moreover, the onset of convective mode of heat transfer (and its increasing impact) with increase in the inclination angle can be clearly revealed from the mid-plane stream function profiles (see Fig. 10). Furthermore, at a given non zero inclination angle, with increase the absorption coefficient, the magnitude of the stream function also increases.

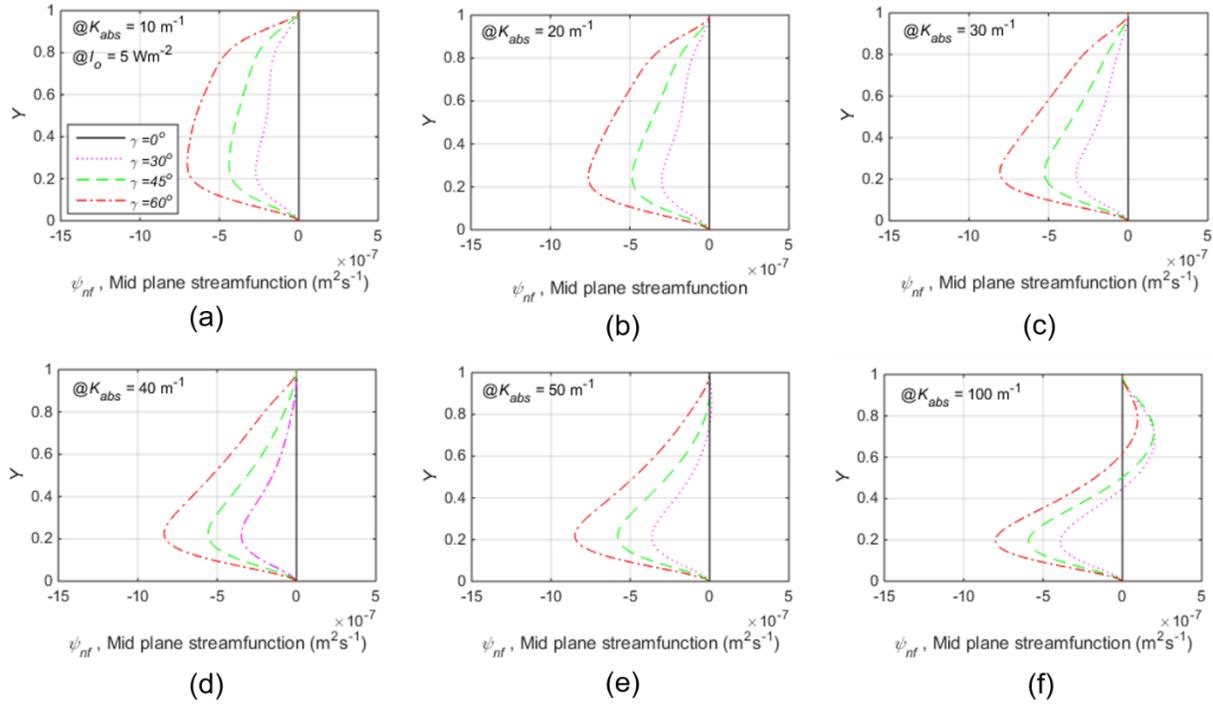

Fig. 10 Mid-plane stream function profiles for volumetric absorption nanofluid based enclosures under different inclination angles, $I_o$ = 5 Wm$^{-2}$ and absorption coefficient (a) $K_{abs}$ = 10 m$^{-1}$ (b) $K_{abs}$ = 20 m$^{-1}$ (c) $K_{abs}$ = 30 m$^{-1}$ (d) $K_{abs}$ = 40 m$^{-1}$ (e) $K_{abs}$ = 50 m$^{-1}$ (f) $K_{abs}$ = 100 m$^{-1}$.

To consider the effect of incident flux; $I_o$ = 50 Wm$^{-2}$ has also been analyzed under similar operating conditions. Stream function plots show trends similar to that for $I_o$ = 5 Wm$^{-2}$ under all considered angles and absorption coefficients. The only difference being that due to increased convection effects, the magnitude of stream function values is higher (see Fig. 11) - hence more uniform temperature distribution.

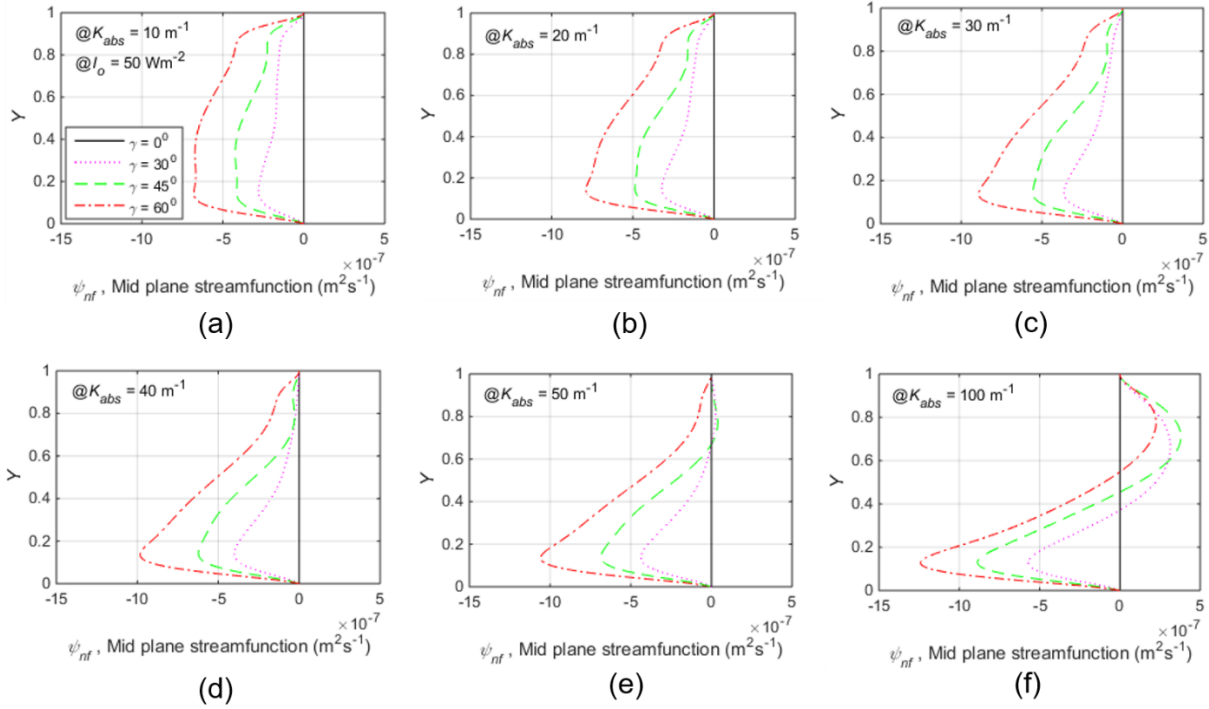

Fig. 11 Mid-plane stream function profiles for volumetric absorption nanofluid filled enclosures under different inclination angles, $I_o = 50$ Wm$^{-2}$ and absorption coefficient (a) $K_{abs} = 10$ m$^{-1}$ (b) $K_{abs} = 20$ m$^{-1}$ (c) $K_{abs} = 30$ m$^{-1}$ (d) $K_{abs} = 40$ m$^{-1}$ (e) $K_{abs} = 50$ m$^{-1}$ (f) $K_{abs} = 100$ m$^{-1}$.

Overall, in 'mixed to volumetric' absorption mode, bottom irradiated volumetric absorption nanofluid filled enclosure at $\gamma = 0^o$, shows conduction dominated heat flow at extreme low absorption coefficient, which changes to conduction plus heat generation phenomenon as absorption coefficient approaches higher values. Finally, convective flows start dominating as the enclosure angle increases and make a maximum effect at an extreme angle, $\gamma = 60^o$, $I_o = 50$ Wm$^{-2}$, resulting in more uniform temperature field.

### 4.2.2 'Surface' absorption mode (adiabatic boundaries)

At very high values of absorption coefficient ($K_{abs} = 100000$ m$^{-1}$), all the incident flux is absorbed in the few bottom layers in the enclosure. The energy redistribution thereafter predominantly happens via convection - typically simulating the characteristic surface absorption RBC.

*Effect of incident flux and enclosure inclination angle:* Figures 12 and 13 show the temperature field (for extreme cases, $I_o = 5$ Wm$^{-2}$ and $I_o = 50$ Wm$^{-2}$) and various inclination angles at steady state. Unequivocally, in all the cases, uniform temperature (throughout the cavity) is obtained irrespective of the incident flux or inclination angle.

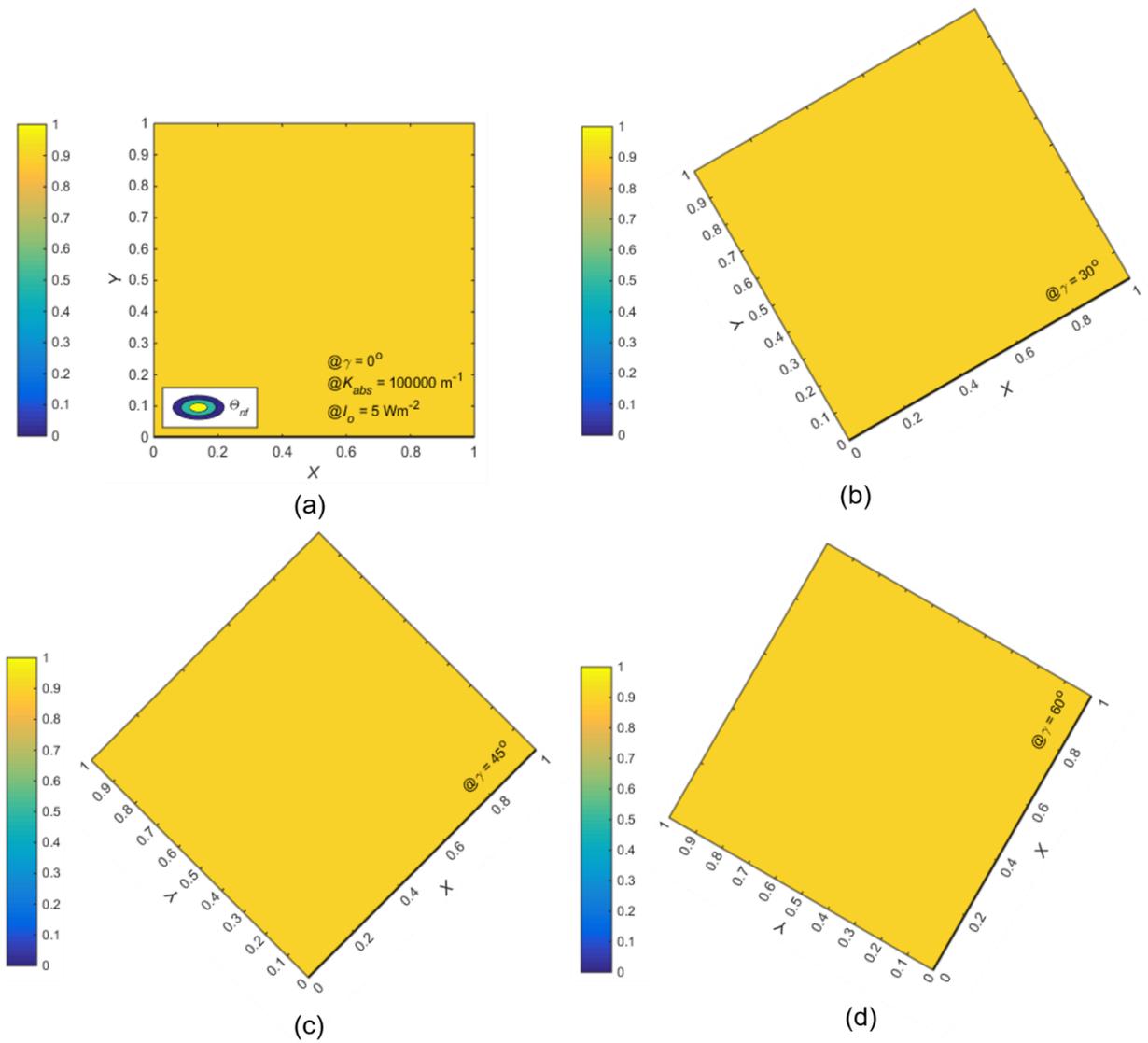

Fig. 12 Temperature contours for 'surface' absorption mode of nanofluid filled enclosures for incident flux, $I_o$ = 5 Wm$^{-2}$, and absorption coefficient, $K_{abs}$ = 100000 m$^{-1}$ at (a) $\gamma = 0°$, (b) $\gamma = 30°$, (c) $\gamma = 45°$, and (d) $\gamma = 60°$.

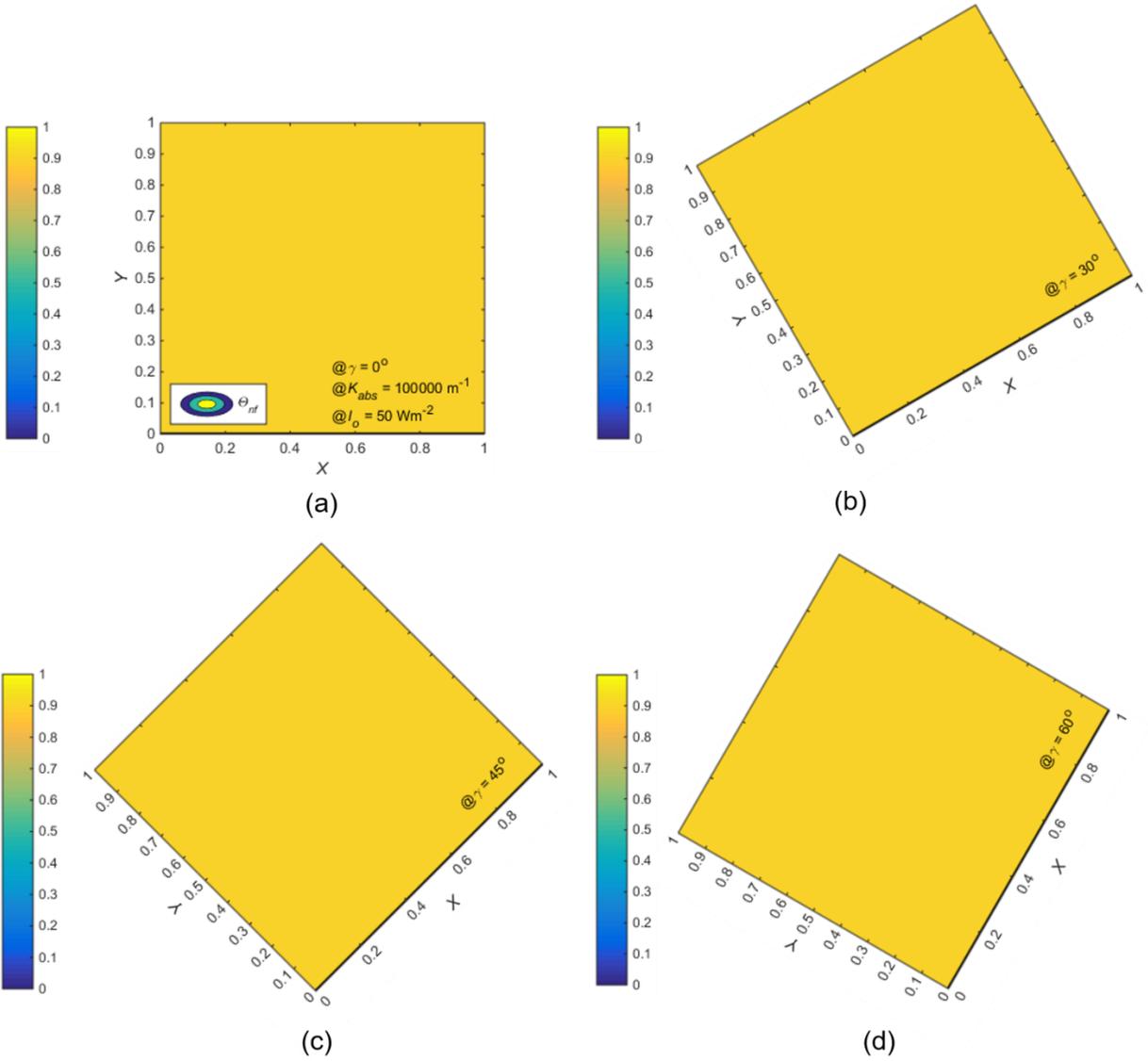

Fig. 13 Temperature contours for 'surface' absorption mode of nanofluid filled enclosures for incident flux, $I_o$ = 50 Wm$^{-2}$, and absorption coefficient, $K_{abs}$ = 100000 m$^{-1}$ at (a) $\gamma = 0°$, (b) $\gamma = 30°$, (c) $\gamma = 45°$, and (d) $\gamma = 60°$.

A closer look into the evolution of temperature field (from start to steady state, see Fig. 14) clearly points out that after initial conduction dominated transport mechanisms, there is onset of convection, which strengthens as we move forward in time and reaches a maxima. Further movement in time is characterized by weakening of convection strength and finally, at steady state, uniform temperature (negligibly small temperature gradients) is established in the enclosure. The aforementioned evolution is also confirmed if we delve into details of stream function magnitudes through evolution in time (see Fig. 15).

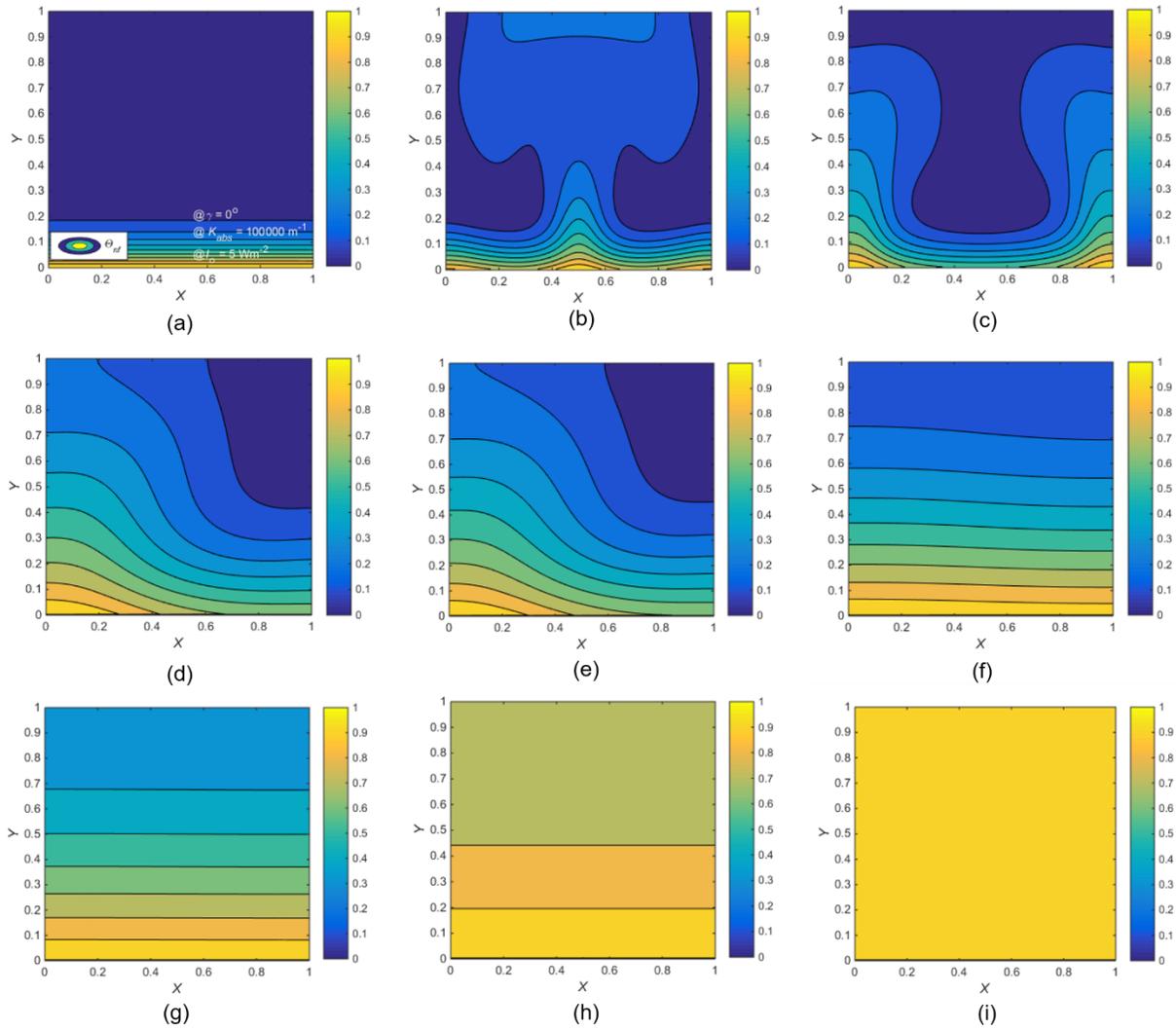

Fig. 14 Variation of temperature distribution for 'surface' absorption mode of nanofluid filled enclosure (viz. $\gamma = 0^o$ and $I_o = 50$ Wm$^{-2}$) until steady state achieved corresponding to time (a) 100 s (b) 5000 s (c) 10000 s (d) 9.2198e+04 s (e) 9.6271e+04 s (f) 1.2177e+05 s (g) 1.2895e+05 (h) 1.5104e+05 s (i) 1.8368e+05 s

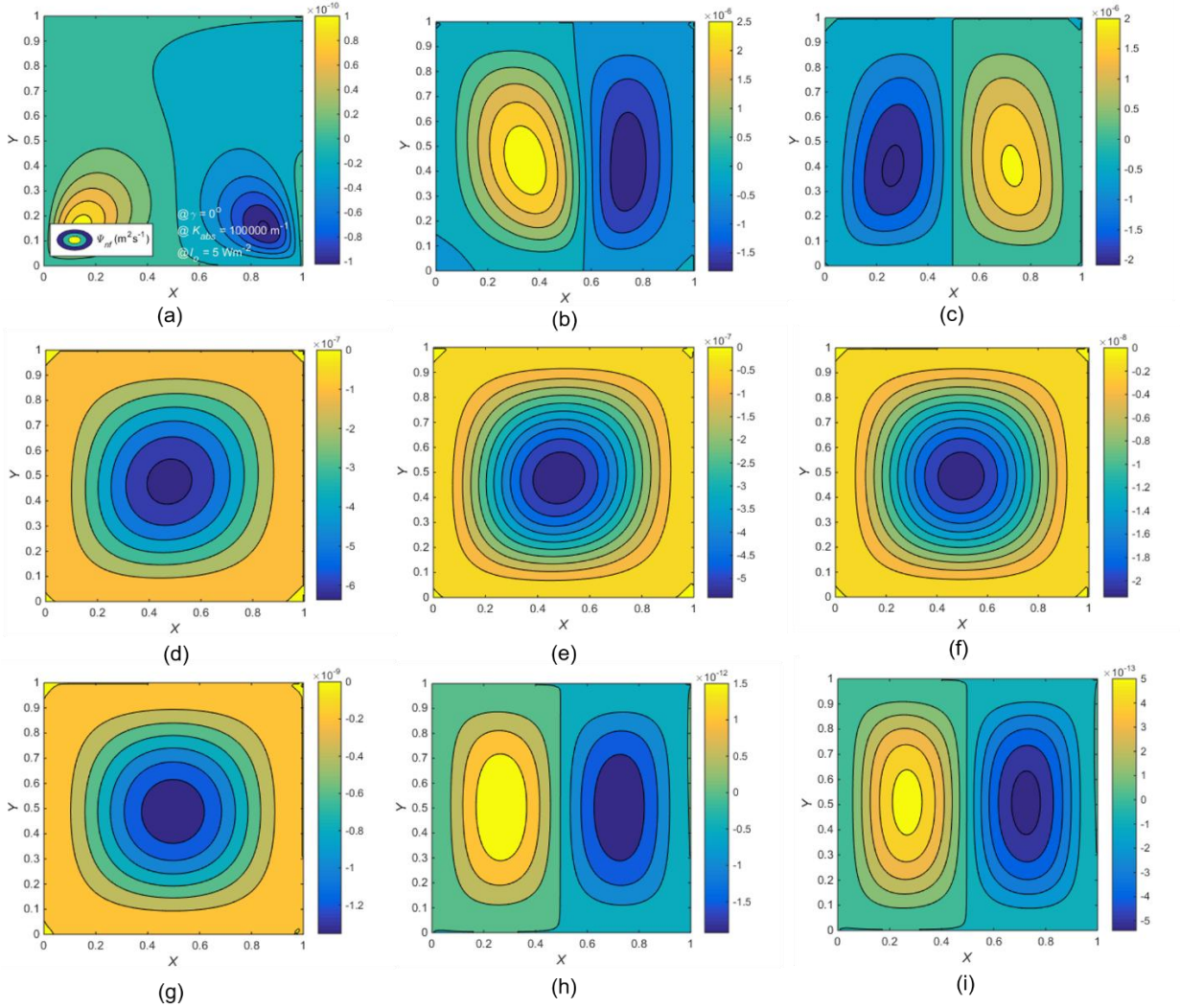

Fig. 15 Variation of stream function for 'surface' absorption mode of nanofluid filled enclosure (viz. $\gamma = 0^o$ and $I_o = 50$ Wm$^{-2}$) until steady state achieved corresponding to time (a) 100 s (b) 5000 s (c) 10000 s (d) 9.2198e+04 s (e) 9.6271e+04 s (f) 1.2177e+05 s (g) 1.2895e+05 (h) 1.5104e+05 s (i) 1.8368e+05 s

### 4.3 Assessing the key transport mechanisms in 'surface' and 'mixed to volumetric' absorption modes in nanofluid filled enclosure with isothermal boundaries

*4.3.1 'Mixed to volumetric' absorption mode (isothermal boundaries)*

***Effect of absorption coefficient and incident flux:*** Under isothermal boundaries ($T_L = T_R = T_T = $ 293K), temperature fields in 'surface to volumetric' absorption mode are presented in Fig. 16, (corresponding to $I_o = 5$ Wm$^{-2}$ and $\gamma = 0^o$). Opposed to the corresponding case of 'adiabatic boundaries'; there is pronounced convection in 'the case of 'isothermal boundaries'. Moreover, as the absorption coefficient increases (from $K_{abs} = 10$ m$^{-1}$ to 100 m$^{-1}$), the strength of convection increases. Similar trends are also observed for $I_o = 50$ Wm$^{-2}$, under similar operating conditions;

with even more pronounced convection, resulting in better mixing and hence more uniform temperature field (see Fig. 17). Moreover, the corresponding flow fields (presented in the form of stream functions, see Figs. 18 and 19) also confirm the aforementioned observations.

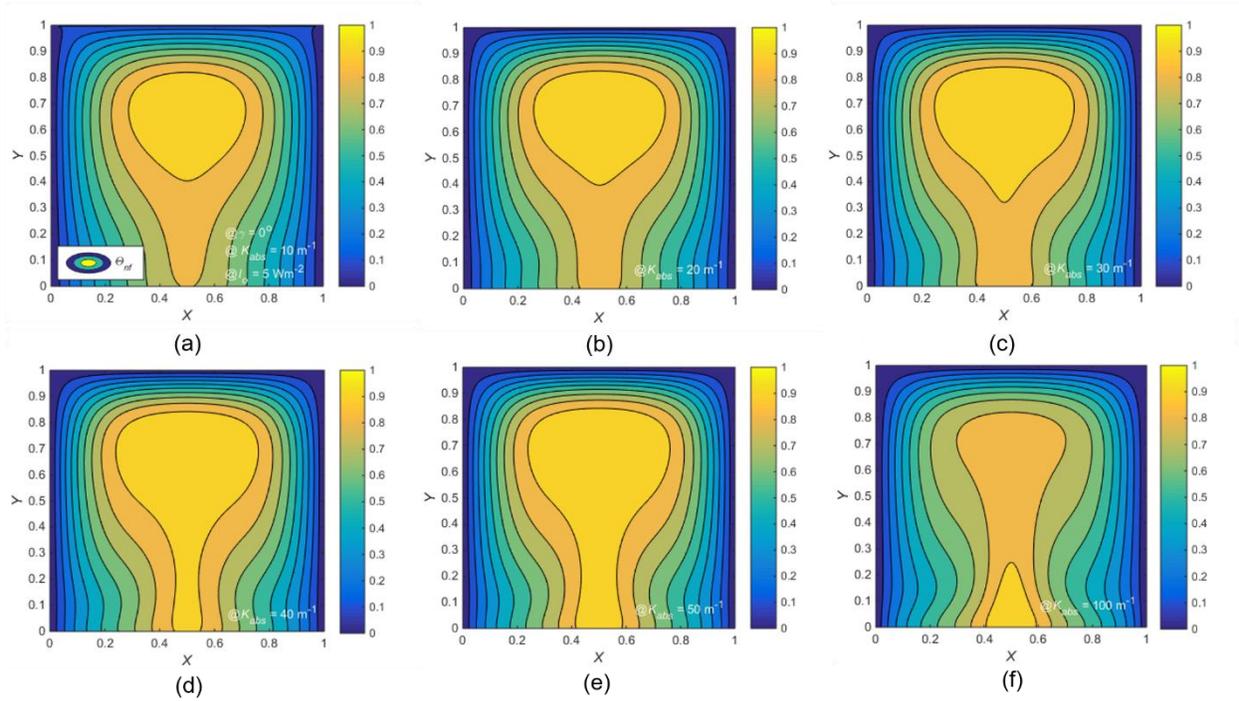

Fig. 16 Temperature contours for 'mixed to volumetric' absorption mode of nanofluid filled enclosures at $\gamma = 0^o$, $I_o = 5$ Wm$^{-2}$ and absorption coefficient (a) $K_{abs} = 10$ m$^{-1}$ (b) $K_{abs} = 20$ m$^{-1}$ (c) $K_{abs} = 30$ m$^{-1}$ (d) $K_{abs} = 40$ m$^{-1}$ (e) $K_{abs} = 50$ m$^{-1}$ (f) $K_{abs} = 100$ m$^{-1}$.

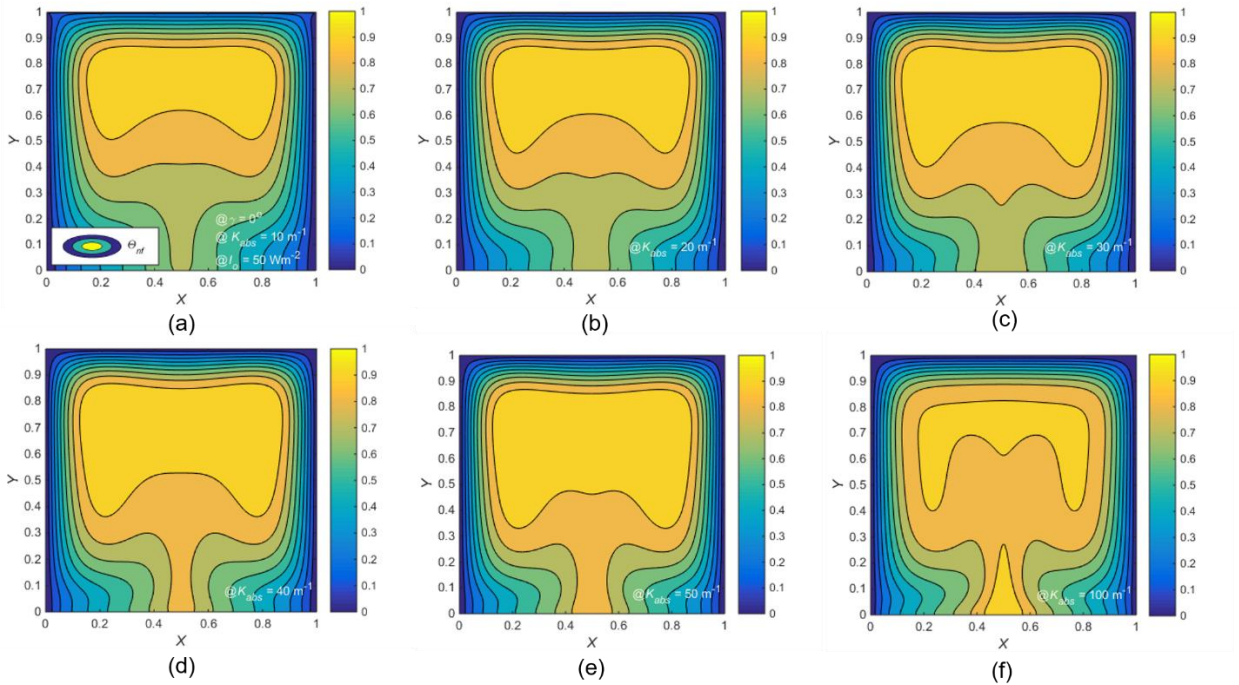

Fig. 17 Temperature contours for 'mixed to volumetric' absorption mode of nanofluid based enclosures at $\gamma = 0°$, $I_o = 50$ Wm$^{-2}$ and absorption coefficient (a) $K_{abs} = 10$ m$^{-1}$ (b) $K_{abs} = 20$ m$^{-1}$ (c) $K_{abs} = 30$ m$^{-1}$ (d) $K_{abs} = 40$ m$^{-1}$ (e) $K_{abs} = 50$ m$^{-1}$ (f) $K_{abs} = 100$ m$^{-1}$.

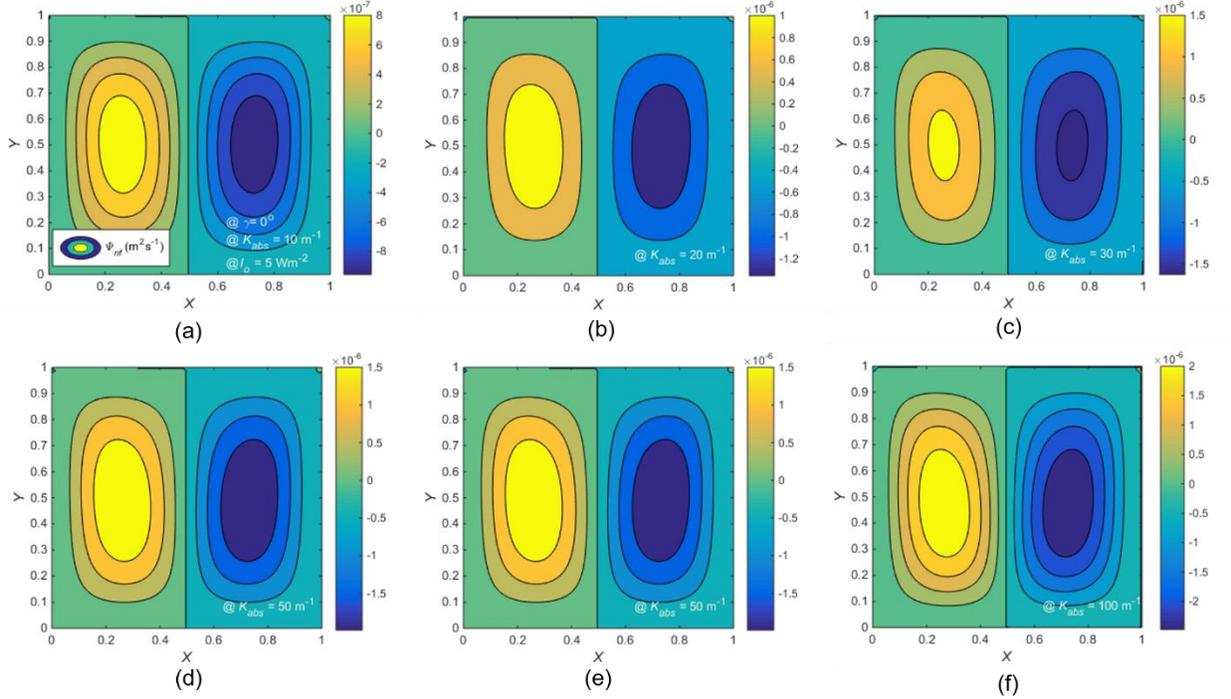

Fig. 18 Stream function contours for 'mixed to volumetric' absorption mode of nanofluid filled enclosures under $\gamma = 0°$, $I_o = 5$ Wm$^{-2}$ and absorption coefficient (a) $K_{abs} = 10$ m$^{-1}$ (b) $K_{abs} = 20$ m$^{-1}$ (c) $K_{abs} = 30$ m$^{-1}$ (d) $K_{abs} = 40$ m$^{-1}$ (e) $Kabs = 50$ m$^{-1}$ (f) $K_{abs} = 100$ m$^{-1}$.

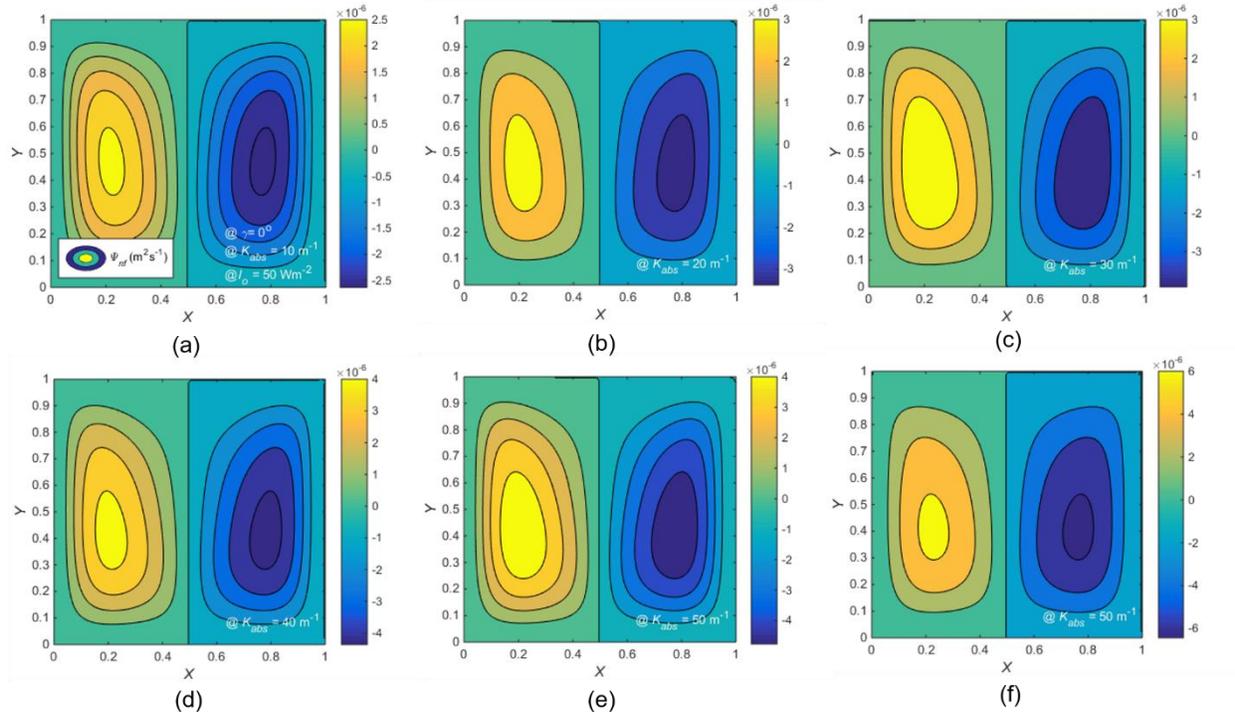

Fig. 19 Stream function contours for 'mixed to volumetric' absorption mode of nanofluid filled enclosures under $\gamma = 0^o$, $I_o = 50$ Wm$^{-2}$ and absorption coefficient (a) $K_{abs} = 10$ m$^{-1}$ (b) $K_{abs} = 20$ m$^{-1}$ (c) $K_{abs} = 30$ m$^{-1}$ (d) $K_{abs} = 40$ m$^{-1}$ (e) $K_{abs} = 50$ m$^{-1}$ (f) $K_{abs} = 100$ m$^{-1}$.

*Effect of enclosure inclination angle and incident flux:* Enclosure inclination angles other than $\gamma = 0^o$ are characterized by strong convection currents (increasing strength from $\gamma = 30^o$ towards $\gamma = 60^o$). Herein, convective flow starts with stronger positive rolls and weaker negative rolls, which are subsequently taken over completely by positive rolls as absorption coefficient approaches high value (i.e., $K_{abs} = 100$ m$^{-1}$). Moreover, the convection strength is directly proportional to incident flux. Figures 20, 21 and 22 show the stream function trends for $\gamma = 30^o$, $45^o$, and $60^o$ respectively.

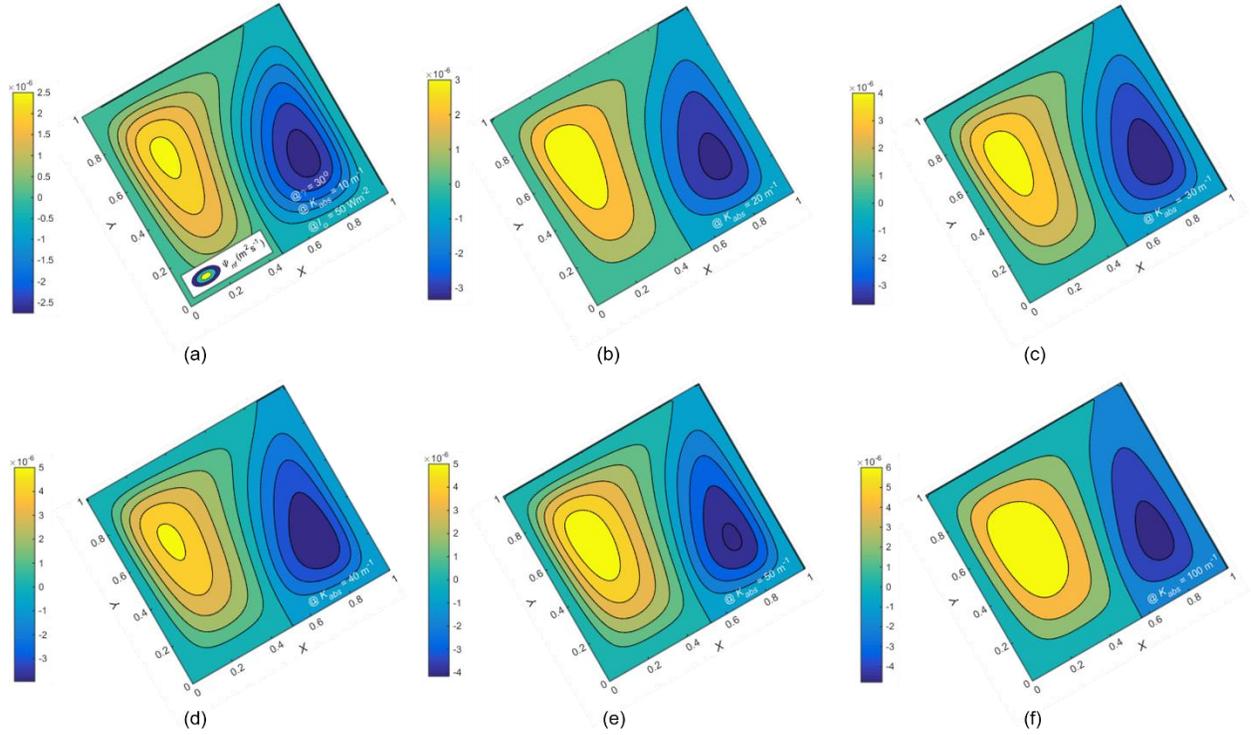

Fig. 20 Stream function contours for 'mixed to volumetric' absorption mode of nanofluid filled enclosures under $\gamma = 30°$, $I_o = 50$ Wm$^{-2}$ and absorption coefficient (a) $K_{abs} = 10$ m$^{-1}$ (b) $K_{abs} = 20$ m$^{-1}$ (c) $K_{abs} = 30$ m$^{-1}$ (d) $K_{abs} = 40$ m$^{-1}$ (e) $K_{abs} = 50$ m$^{-1}$ (f) $K_{abs} = 100$ m$^{-1}$.

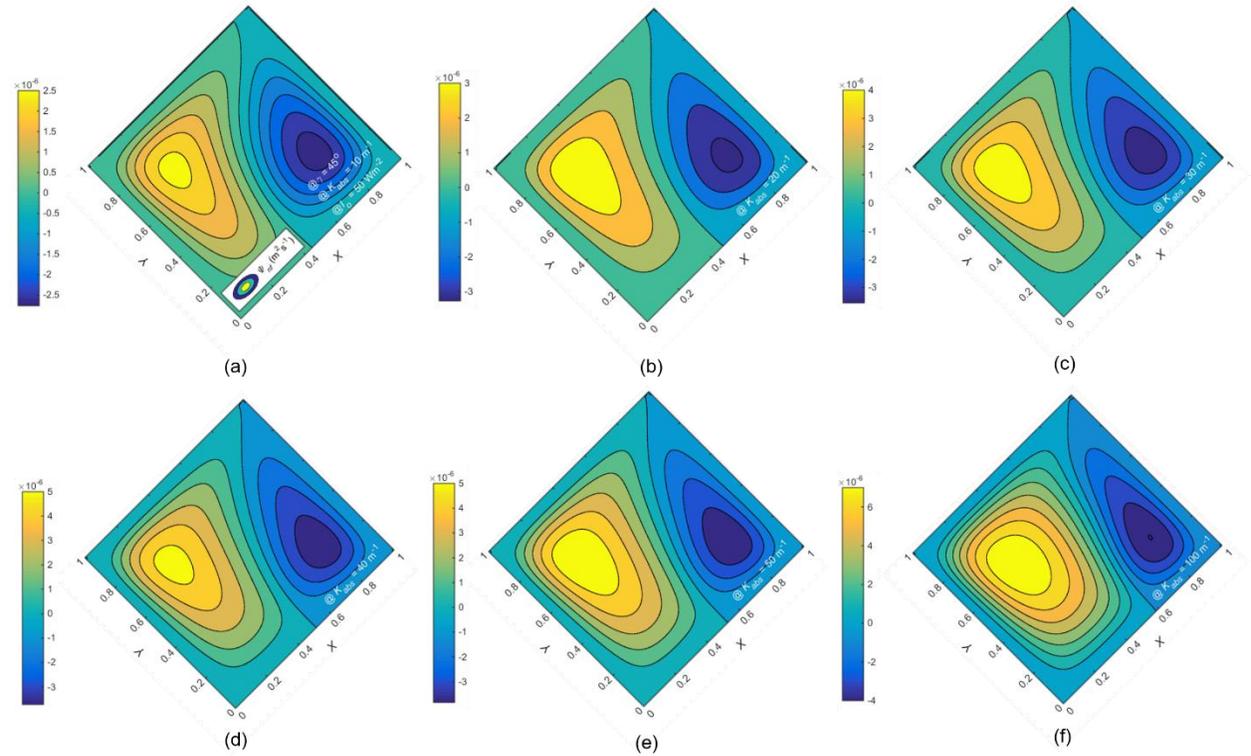

Fig. 21 Stream function contours for 'mixed to volumetric' absorption mode of nanofluid filled enclosures under $\gamma = 45^{\circ}$, $I_o = 50$ Wm$^{-2}$ and absorption coefficient (a) $K_{abs} = 10$ m$^{-1}$ (b) $K_{abs} = 20$ m$^{-1}$ (c) $K_{abs} = 30$ m$^{-1}$ (d) $K_{abs} = 40$ m$^{-1}$ (e) $K_{abs} = 50$ m$^{-1}$ (f) $K_{abs} = 100$ m$^{-1}$.

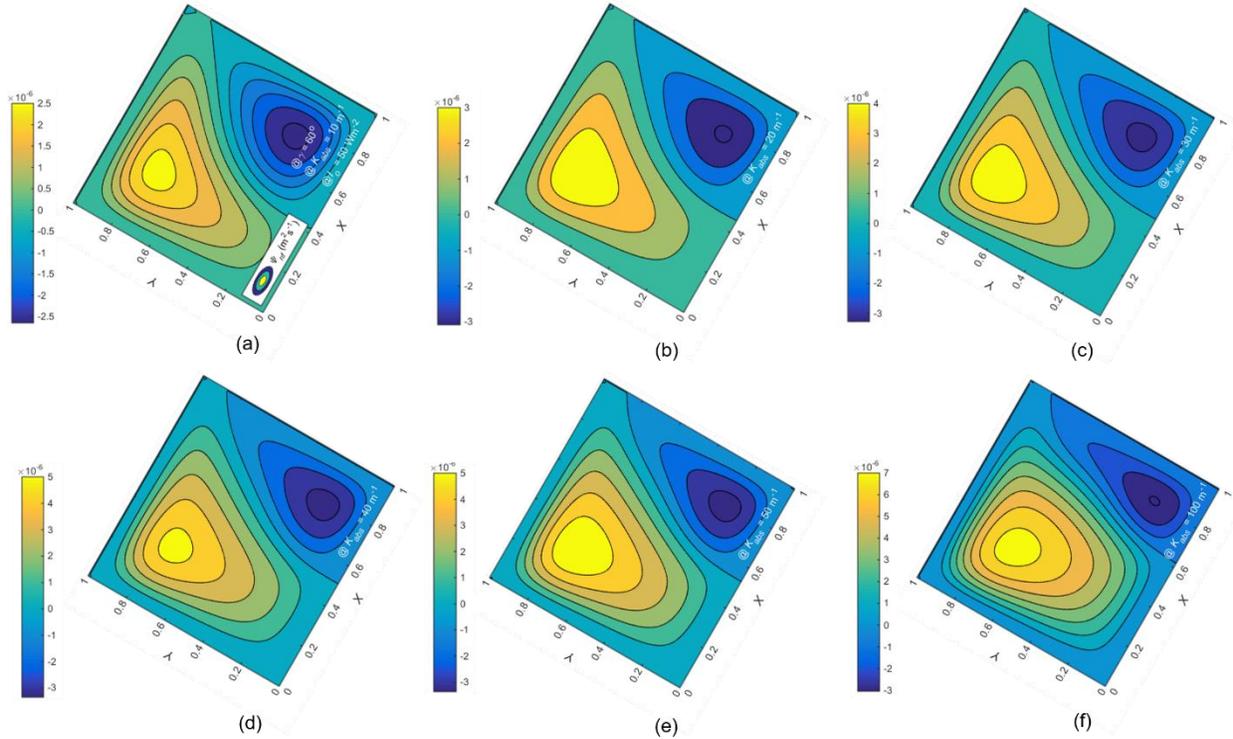

Fig. 22 Stream function contours for 'mixed to volumetric' absorption mode of nanofluid filled enclosures under $\gamma = 60^{\circ}$, $I_o = 50$ Wm$^{-2}$ and absorption coefficient (a) $K_{abs} = 10$ m$^{-1}$ (b) $K_{abs} = 20$ m$^{-1}$ (c) $K_{abs} = 30$ m$^{-1}$ (d) $K_{abs} = 40$ m$^{-1}$ (e) $K_{abs} = 50$ m$^{-1}$ (f) $K_{abs} = 100$ m$^{-1}$.

### *4.3.2 'Surface' absorption mode (isothermal boundaries)*

*Effect of incident flux:* Focusing on 'surface' absorption mode nanofluid-filled enclosure, we see convection typically RBC. Figure 23 shows temperature contours at $K_{abs} = 100000$ m$^{-1}$flux, for different incident flux values for $\gamma = 0^{\circ}$. Due to isothermal boundaries, counter rotating convective rolls ensure the symmetrical distribution of temperature about the mid-plane, with generation of plumes from bottom center to near top boundary of the enclosure. Clearly, the plume becomes sharper as the value of the incident flux increases - depicting increasing convection strength with increasing flux.

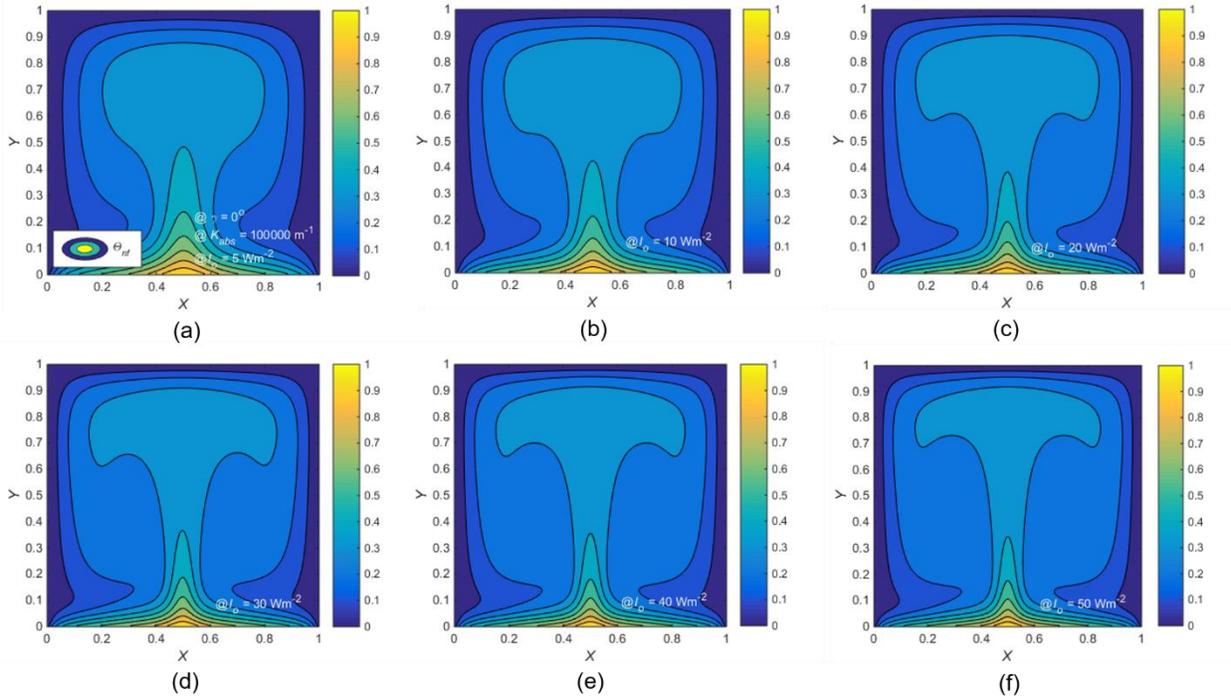

Fig. 23 Temperature contours for 'surface' absorption mode of nanofluid based enclosures (depicting RBC) at $K_{abs}$ = 100000 m$^{-1}$, under fluxes (a) $I_o$ = 5 Wm$^{-2}$ (b) $I_o$ = 10 Wm$^{-2}$ (c) $I_o$ = 20 Wm$^{-2}$ (d) $I_o$ = 30 Wm$^{-2}$ (e) $I_o$ = 40 Wm$^{-2}$ (f) $I_o$ = 50 Wm$^{-2}$

*Effect of incident flux and enclosure inclination angle:* Furthermore, effect of inclination angle with varying flux is investigated via mid plane temperature profiles (see Fig. 24). Unlike the case of enclosure inclination $\gamma$ = 0 °; unconditional steady solutions are not observed for other enclosure inclinations (i.e., $\gamma$ = 30 °, 40 °, and 60 °). In other words, the solutions obtained for all inclination angles except ($\gamma$ = 0 °) are not converging - signifying transition regime. Interestingly, the incident flux value at which transition happens is also characteristic of the enclosure inclination. At the enclosure inclination angle of $\gamma$ = 30 °, the solution starts oscillating from flux, $I_o$ = 20 Wm$^{-2}$ onwards. Furthermore, the solution starts oscillating for flux $I_o$ = 30 Wm$^{-2}$ onwards at an inclination angle, $\gamma$ = 45 °. Finally, corresponding to $\gamma$ = 60 °, solution give oscillations for flux *($I_o$ = 40 Wm$^{-2}$ onwards).*To understand the oscillations, which repeat after certain time intervals; results (of a representative case) corresponding to $\gamma$ = 30 ° and $I_o$ = 20 Wm$^{-2}$ depicting Rayleigh Benard convection are presented in Fig. 25 at different points in time.

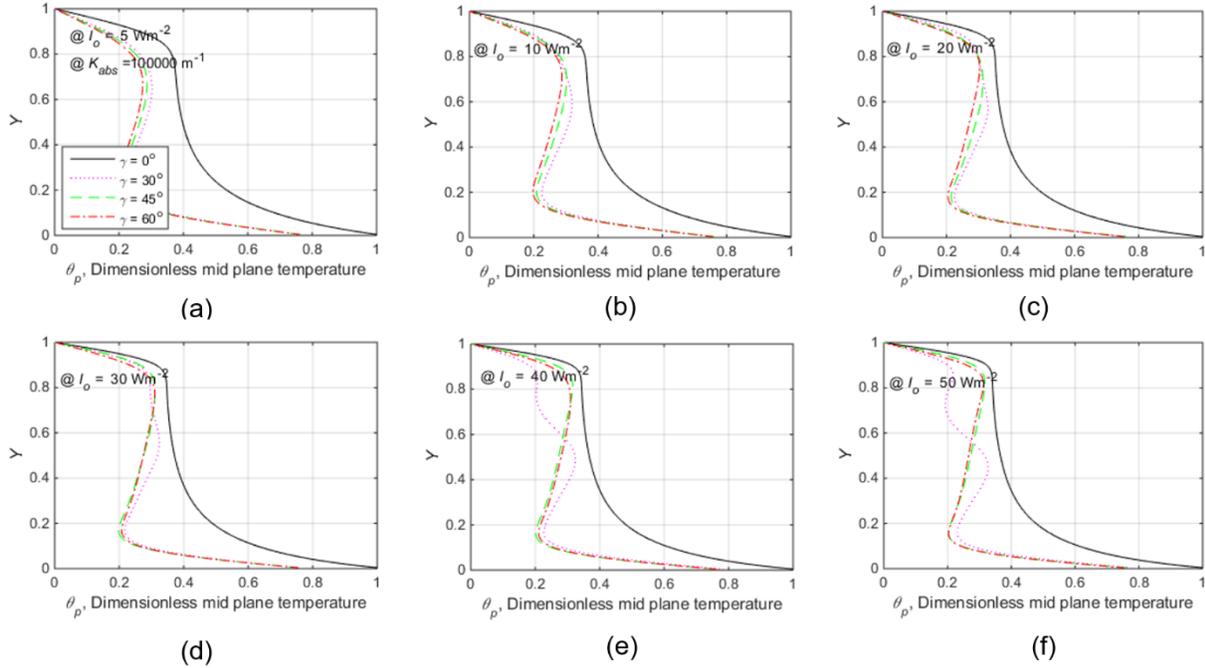

Fig. 24 Mid-plane temperature profiles for 'surface' absorption mode of nanofluid filled enclosures under different fluxes, and inclination angles (a) $\gamma = 0^o$ (b) $\gamma = 30^o$ (c) $\gamma = 45^o$ (d) $\gamma = 60^o$.

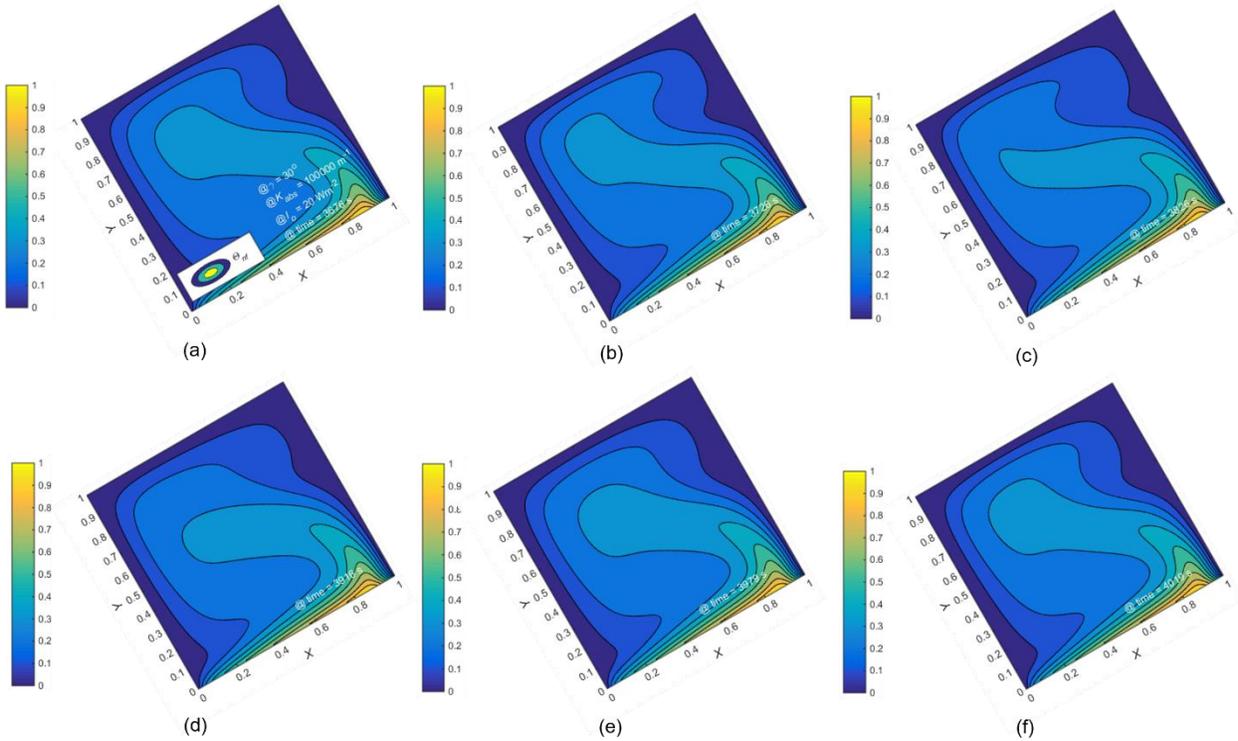

Fig. 25 Temperature contours during one period of oscillation corresponding to $\gamma = 30^o$, $I_o = 20$ Wm$^{-2}$, and $K_{abs} = 100000$ m$^{-1}$ under isothermal boundaries.

## CONCLUSIONS

Relative importance of different heat transport mechanisms in nanofluid filled enclosures (irradiated from below) in "volumetric", "mixed" and "surface" absorption modes has been quantified. Herein, the prime focus is to understand the impact of enclosure inclination angle, incident flux and boundary conditions on the temperature and flow field. Considering adiabatic boundaries (for $\gamma = 0^o$) with "mixed to volumetric" absorption mode, heat transport invariably starts with pure conduction (at extreme low absorption coefficients) to "conduction plus generation" (at higher absorption coefficients), and always remain free of natural convection effects. However for non-zero values of enclosure inclination angle, convection currents are always present. Furthermore, in "surface" absorption mode, RBC types of flows are the norm irrespective of the incident flux values and enclosure inclination angles.

For isothermal boundary condition, convective flows are observed irrespective of the nature of the absorption mode. Moreover, for non-zero enclosure inclination angles, even transition to transient state is observed. Overall, the present work is significant step towards deciphering the predominant transport mechanisms operational in bottom irradiated nanofluid filled enclosures- thus helping in understanding numerous radiative driven flows in nature and industrial processes.

## ACKNOWLEDGEMENTS

IS and SS acknowledge the support provided by Mechanical Engineering Department, Chandigarh University, Gharuan. V.K. acknowledges the support provided by Mechanical Engineering Department, Thapar Institute of Engineering & Technology Patiala, India.

## NOMENCLATURE

**English Symbols:**

| Symbol | Description | Symbol | Description |
|---|---|---|---|
| $a$ | surface area [m$^2$] | $\mu$ | dynamic viscosity [kgm$^{-1}$s$^{-1}$] |
| $C_{ps}$ | specific heat of nanofluid [Jkg$^{-1}$K$^{-1}$] | $\vartheta$ | kinematic viscosity of nanofluid [m$^2$s$^{-1}$] |
| $D$ | width of enclosure [m] | $\Psi$ | stream function [m$^2$s$^{-1}$] |
| $g$ | gravitational effect [ms$^{-2}$] | $\psi$ | mid plane stream function [m$^2$s$^{-1}$] |
| $H$ | height of enclosure [m] | $\Theta$ | dimensionless local fluid temperature, $(T_{nf}-T_{min})/(T_{max}-T_{min})$ |
| $h$ | convection heat transfer coefficient [Wm$^{-2}$K$^{-1}$] | $\theta$ | dimensionless mid plane temperature, $(T_p-T_{min})/(T_{max}-T_{min})$ |
| $I_o$ | incident flux [Wm$^{-2}$] | $\rho$ | density of nanofluid |
| $I$ | local flux [Wm$^{-2}$] | $\sigma$ | stefan boltzmann constant [=5.67×10$^{-8}$ Wm$^{-2}$K$^{-4}$] |
| $K_{abs}$ | absorption coefficient [m$^{-1}$] | $\tau$ | glass transmissivity |
| $k$ | thermal conductivity [Wm$^{-1}$K$^{-1}$] | $\varepsilon$ | emissivity of glass plate |
| $L$ | length of enclosure [m] | **Subscript:** | |
| $Pr$ | Prandtl Number | $abs$ | absorption |
| $R^o$ | dimensionless flux [$I/I_o$] | $amb$ | ambient |
| $S^T$ | radiative source term [Wm$^{-1}$] | $B$ | bottom wall |
| $T$ | local fluid temperature [K] | $L$ | left wall |
| $T_p$ | mid plane temperature [K] | $nf$ | Nanofluid |
| $u$ | velocity component of fluid along x- | $R$ | right wall |
|   | direction [ms$^{-1}$] |   |   |
| *v* | velocity component of fluid along y-direction [ms$^{-1}$] | *r* | radiation |
| *x* | horizontal coordinate direction | *ref* | reference |
| *y* | vertical coordinate direction | *s* | surface |
| *X* | dimensionless enclosure width, *x/D* | *T* | top wall |
| *Y* | dimensionless enclosure height, *y/H* | **Superscript:** | |
| **Greek symbols:** | | T | Term |
| α | thermal diffusivity [m$^2$s$^{-1}$] | **Abbreviation** | |
| *β* | coefficient of volumetric expansion [K$^{-1}$] | RBC | Rayleigh Benard convection |